
\magnification\magstep1
\font\BBig=cmr10 scaled\magstep2
\font\BBBig=cmr10 scaled\magstep3


\def\title{
{\bf\BBBig
\centerline{Conformal symmetry}
\bigskip
\centerline{of the coupled Chern-Simons and
}\bigskip
\centerline{gauged nonlinear Schr\"odinger equations}
}
} 


\def\authors{
\centerline{
C.~DUVAL\foot{
D\'epartement de Physique, Universit\'e
d'Aix-Marseille II and
Centre de Physique
Th\'eorique CNRS, Luminy Case 907, F--13288 MARSEILLE,
Cedex 09 (France);
e-mail: duval@marcptsu1.univ-mrs.fr},
P.~A.~HORV\'ATHY\foot{
D\'epartement de Math\'ematiques,
Universit\'e de Tours, Parc de Grandmont,
F--37200 TOURS (France);
e-mail: horvathy@univ-tours.fr
}
and
L.~PALLA\foot{
Institute for Theoretical Physics,
E\"otv\"os University, H--1088 BUDAPEST,
Puskin u. 5--7 (Hungary);
e-mail: palla@ludens.elte.hu
}
}
}

\def\runningauthors{
Duval, Horv\'athy \& Palla
}

\def\runningtitle{
Chern-Simons and gauged non-linear Schr\"odinger\dots
}


\voffset = 1cm 
\baselineskip = 14pt 

\headline ={
\ifnum\pageno=1\hfill
\else\ifodd\pageno\hfil\tenit\runningtitle\hfil\tenrm\folio
\else\tenrm\folio\hfil\tenit\runningauthors\hfil
\fi
\fi}

\nopagenumbers
\footline={\hfil} 


\font\tenb=cmmib10
\newfam\bsfam

\textfont\bsfam=\tenb

\mathchardef\betab="080C
\mathchardef\xib="0818
\mathchardef\omegab="0821
\mathchardef\deltab="080E
\mathchardef\epsilonb="080F
\mathchardef\pib="0819
\mathchardef\sigmab="081B
\mathchardef\bfalpha="080B
\mathchardef\bfbeta="080C
\mathchardef\bfgamma="080D
\mathchardef\bfomega="0821
\mathchardef\zetab="0810
\mathchardef\gammab="080D


\def\and{\qquad\hbox{and}\qquad}

\def\kikezd{\parag\underbar} 

\def\IR{{\bf R}}
\def\smallover#1/#2{\hbox{$\textstyle{#1\over#2}$}}
\def\2{{\smallover 1/2}}
\def\ccr{\cr\noalign{\medskip}} 
\def\parag{\hfil\break} 
\def\={\!=\!}
\def\boxit#1{
\vbox{\hrule\hbox{\vrule\kern3pt
\vbox{\kern3pt#1\kern3pt}\kern3pt\vrule}\hrule}
} 

\def\dAlembert{\boxit{\null}\,} 


\newcount\ch 
\newcount\eq 
\newcount\foo 
\newcount\ref 

\def\chapter#1{
\parag\eq = 1\advance\ch by 1{\bf\the\ch.\enskip#1}
}

\def\equation{
\leqno(\the\ch.\the\eq)\global\advance\eq by 1
}

\def\foot#1{
\footnote{($^{\the\foo}$)}{#1}\advance\foo by 1
} 

\def\reference{
\parag [\number\ref]\ \advance\ref by 1
}

\ch = 0 
\foo = 1 
\ref = 1 


\title
\vskip1.5cm
\authors
\vskip.5in

\parag
{\bf Abstract.}

{\it The non-relativistic conformal symmetry found by
Jackiw and Pi for the coupled Chern-Simons and
gauged nonlinear Schr\"odinger equations in the plane is derived
in a non-relativistic Kaluza-Klein framework.}

\vskip2in
\smallskip
\vfill
\eject

\chapter{Introduction}

Recently, Jackiw and Pi~[1] found that the
{\it gauged, planar non-linear Schr\"odinger equation}
$$
i\partial_t\Psi=\left\{
-{1\over2m}(\vec{\nabla}-ie\vec{A})^2+eA^0
-\Lambda\,\Psi^*\Psi\right\}\Psi,
\equation
$$
where $(A^0,\vec{A})$ is an electromagnetic vector potential
and $\Lambda={\rm const}.$,
admits a `non-relativistic conformal' symmetry~[2].
The electromagnetic field
and the current $(\rho,\vec{J})$,
$$
\rho=\Psi^*\Psi,
\qquad
\vec{J}={1\over 2im}\left[
\Psi^*\vec{D}\Psi-\Psi(\vec{D}\Psi)^*
\right],
\equation
$$
are assumed to satisfy the field-current identity
$$
B=\epsilon^{ij}\partial_iA^j=-{e\over\kappa}\rho,
\qquad
E^i=-\partial_iA^0-\partial_tA^i={e\over\kappa}\,\epsilon^{ij}J^j,
\equation
$$
($i,j=1,2$), rather than the conventional Maxwell equations. For a
special value of $\Lambda$ these equations admit static solutions~[1],
interpreted as Chern-Simons vortices.

In this Letter, we investigate non-relativistic Chern-Simons
theory in a Kaluza-Klein-type framework.
The clue is that non-relativistic space-time in
$2+1$ dimensions is conveniently
viewed as the quotient of a $(3+1)$-dimensional Lorentz manifold
by the integral curves of a
covariantly constant, lightlike vector field~[3].
Then the non-relativistic conformal
symmetries found by Jackiw and Pi are deduced
from those --- relativistic --- which arise in extended space.

The peculiar aspect of our approach is that
we mainly work with the field equations,
since we have not yet been able to lift completely the
$3$-dimensional variational approach~[1] to $4$ dimensions.

Although our theory is formulated in full generality,
the only example we present here
is the one in Ref.~[1], reproduced by reduction
from Minkowski space. However, the Chern-Simons vortices
in external harmonic and uniform magnetic fields found in Ref.~[4]
fit also into our framework. This will be explained in a forthcoming
publication.

\chapter{Chern-Simons theory in Bargmann space}

Let $(M,g)$ denote a Lorentz $4$-manifold of signature
$(-,+,+,+)$ which is endowed with a complete,
{\it covariantly constant} and
{\it lightlike} vector field $\xi$. Following
our previous terminology~[3], we call such a manifold
a Bargmann space.
Then the quotient, $Q$, of $M$ by the integral
curves of $\xi$ is a $(2+1)$-dimensional
manifold which carries a Newton-Cartan structure,
i.e. the geometric structure of non-relativistic space-time
and gravitation.
An adapted coordinate system on $M$ is provided by $(t,x,y,s)$
where $(t,x,y)$ are coordinates on $Q$ and
$\xi=\partial_s$.
The couple $(x,y)$ yields position coordinates and
$t$ is non-relativistic absolute time,
$g_{\mu\nu}\xi^\nu=\partial_\mu{t}$.

\goodbreak

Consider first a $U(1)$ vector-potential
$a_\mu dx^\mu$ on $M$, denote by
$f={1\over2}\,f_{\mu\nu}\,dx^\mu\wedge dx^\nu$
its field strength,
$f_{\mu\nu}=2\partial_{[\mu}a_{\nu]}$, and
let $j^\mu$ be some given current on
$M$. Let us now {\it postulate} on $M$ the
{\it field-current identity} (FCI)
$$
\kappa f_{\mu\nu}=
e\sqrt{-g}\,\epsilon_{\mu\nu\rho\sigma}\,\xi^\rho j^\sigma,
\equation
$$
($\mu,\nu,\rho,\sigma=0,\ldots,3$). It follows that
$f_{\mu\nu}\xi^\nu=0$ and since $f$ is closed,
$\partial_{[\mu}f_{\nu\rho]}=0$, $f$ is the lift from the
$(2+1)$-dimensional space-time, $Q$, of a closed $2$-form
$F={1\over2}\,F_{\alpha\beta}\,dx^\alpha\wedge dx^\beta$,
($\alpha,\beta=0,1,2$). With no loss of generality, $(a_\mu)$ can be
chosen to be the lift from $Q$ of a vector potential
$(A_\alpha)=(A_0,\vec{A})$ for $F$. Since $\xi$ is covariantly
constant, Eq.~(2.1) implies also that the Lie derivative
of the current
with respect to $\xi$
is proportional to $\xi$; the four-current
$(j^\mu)$ projects therefore into a three-current,
$(J^\alpha)=(\rho,\vec{J})$, on $Q$.
Observe, finally, that the determinant $g=\det(g_{\mu\nu})$ is
$\xi$-invariant so that the contraction of the four-volume element by
the covariantly constant vector $\xi$, namely
$\sqrt{-g}\,\epsilon_{\mu\nu\rho\sigma}\,\xi^\sigma$,
projects to a preferred volume element on Galilean space-time, $Q$.
Therefore, Eq.~(2.1) itself descends to the quotient to yield
$$
\kappa F_{\alpha\beta}
=
-e\sqrt{-g}\,\epsilon_{\alpha\beta\gamma}\,J^\gamma,
\equation
$$
($\alpha,\beta,\gamma=0,1,2$), which constitutes the Galilean
version of the Chern-Simons equations. Let, for example, $M$ be
Minkowski space with its flat metric $dx^2+dy^2+2dtds$ where $t$ and
$s$ are light-cone coordinates. Putting $\xi=\partial_s$ we
plainly get a Bargmann structure over Galilean space-time. The only
components of the electromagnetic field are $E^x=F_{xt}$,
$E^y=F_{yt}$ and $B=F_{xy}$, which satisfy
$\vec{\nabla}\times\vec{E}+\partial_tB=0$ since $F$ is closed. Then
the projected FCI~(2.2) is readily seen to reduce to that in Ref.~[1],
Eq.~(1.3). Note that Jackiw and Pi take their field-current identity
from relativistic Chern-Simons theory although their system is
non-relativistic.

Consider next a massless relativistic scalar field $\psi$
on $(M,g)$ coupled to the gauge field $a_\mu$ and
described by the gauged, non-linear wave equation (NLWE)
$$
\Big\{D_\mu D^\mu-{R\over6}
+\lambda\psi^*\psi\Big\}\psi=0,
\equation
$$
where $\lambda={\rm const.}$,
$D_\mu=\nabla_\mu-iea_\mu$ and $\nabla$ is the
metric-covariant derivative;
we have also included $R$, the scalar curvature of $g_{\mu\nu}$,
for later convenience.
The {\it mass}, $m$, is then introduced by requiring that
the matter field satisfy the equivariance condition
$$
\xi^\mu D_\mu\psi=im\psi.
\equation
$$
Due to our choice of gauge,
$\Psi=e^{-ims}\psi$ is thus a function on $Q$ and Eq.~(2.3)
can be expressed as an equation on $Q$.
This equation looks in general quite complicated and is not
reproduced here.
In the absence of a gauge field and for $\lambda=0$
it provides a covariant Schr\"odinger equation on Newton-Cartan
space-time~[5,3].
In the flat case, however, things simplify: the NLWE (2.3)
reduces precisely to the non-linear Schr\"odinger equation (1.1)
with $\Lambda=\lambda/2m$.

Our system of equations becomes {\it coupled} by requiring,
in addition to Eqs~(2.1,3,4), that the current be related to
the scalar field according to
$$
j^\mu={1\over2mi}\Big[\psi^*(D^\mu\psi)-\psi(D^\mu\psi)^*\Big].
\equation
$$
We note that Eq.~(2.3) implies that $j^\mu$ is conserved,
$\nabla_\mu j^\mu=0$.
The system (2.1,3$-$5) projects, in the flat case, to Eqs~(1.1$-$3)
on Galilean space-time.

\goodbreak

\chapter{Non-relativistic conformal symmetries}

A conformal transformation $\varphi:M\to M$ is such that
the pulled-back metric is of the form
$\varphi^\star{g}=\Omega^2{g}$ for some strictly positive function
$\Omega$. Such transformations give rise to the conformal group of
$(M,g)$.
If $\varphi$ is also required to preserve the vertical
vector $\xi$, namely $\varphi^\star\xi=\xi$, then it is easy to prove
that $\Omega$ is a function of time, $t$, alone.
This latter condition entails that $\varphi$ defines a transformation
on the quotient, $Q$. These transformations form a subgroup of the
conformal group we call the generalized Schr\"odinger group.
For example, the conformal transformations of (compactified) Minkowski
space $M$ form the group ${\rm O}(4,2)$; those which also preserve the
lightlike vector $\xi=\partial_s$ yield a $9$-dimensional subgroup
identified with the Schr\"odinger group in $2+1$ dimensions~[6].
The infinitesimal action of this group on Minkowski space $M$ is
given by the vector field
$$
(X^\mu)
=
\pmatrix{ -\chi t^2-\delta\,t-\epsilon \hfill \ccr
R\,\vec{x}
-\left(\2\delta+\chi t\right)\vec{x}
+t\vec{\beta}+\vec{\gamma} \hfill \ccr
\2\chi\,r^2-\vec{\beta}\cdot\vec{x}+\eta \hfill\cr
}
\equation
$$
where $r=|\vec{x}|$ and $R\in{\rm so}(2),\,
\vec{\beta},\vec{\gamma}\in\IR^2,\,
\epsilon,\chi,\delta,\eta\in\IR$,
interpreted as rotation, boost, space translation, time translation,
expansion, dilatation and vertical translation.

Now, let us determine which conformal transformations
act as symmetries. By a `symmetry' we mean
here a transformation which takes a solution of
some system of equations into
another solution of this same system.

Consider first the massless wave equation (2.3) on
extended space-time
$(M,g)$ with $a_\mu$ some given vector potential,
and let $\varphi$ be a conformal transformation,
viz $\varphi^\star{g}=\Omega^2g$.
Expanding the term $D_\mu D^\mu\psi$
and using the transformation law of the scalar curvature,
$
\varphi^\star R=\Omega^{-2}\big[R-
6\Omega^{-1}\nabla_\mu\nabla^\mu\Omega
\big],
$
a term-by-term inspection shows that
the NLWE (2.3) goes into
$$
\Omega^{-3}\Big\{\widetilde{D}_\mu\widetilde{D}^\mu
-{R\over6}
+\lambda\widetilde{\psi}^* \widetilde{\psi}\Big\}
\widetilde{\psi}=0
\equation
$$
upon setting
$$
\widetilde{\psi}=\Omega\,\varphi^\star\psi,
\qquad
\widetilde{a}_\mu=(\varphi^\star a)_\mu
\and
\widetilde{D}_\mu=\nabla_\mu-ie\widetilde{a}_\mu.
\equation
$$
Hence, if $\psi$ and $a_\mu$
solve the NLWE (2.3), the same is true for
$\widetilde{\psi}$ and $\widetilde{a}_\mu$:
the full conformal group is, indeed, a symmetry for the NLWE.
Our result generalizes the statement saying that the ungauged
non-linear Klein-Gordon equation on $n$-dimensional Minkowski
space,
$
\{\dAlembert+\lambda(\psi^*\psi)^p\}\psi=0,
$
is conformally invariant precisely for $p=2/(n-2)$~[7].

Let us now suppose that $a_\mu$ satisfies the FCI (2.1).
This latter is manifestly not invariant under arbitrary conformal
transformations since it involves the vector field $\xi$.
Using $\varphi^\star\sqrt{-g}=\Omega^4\sqrt{-g}$ one proves,
however, that if $\varphi$ is assumed
$\xi$-preserving, then
$\widetilde{f}_{\mu\nu}=(\varphi^\star f)_{\mu\nu}$
satisfies the FCI (2.1) with the new current
$$
\tilde{\jmath}{}^\mu=\Omega^4(\varphi^\star j)^\mu.
\equation
$$
The FCI reduces hence the
symmetry to the Schr\"odinger subgroup.

Remembering that $\Omega$ is thus a function of time only, we see
that the $\xi$-preserving conformal transformations also preserve the
equivariance condition (2.4),
$\widetilde{D}_\xi\widetilde{\psi}=im\widetilde{\psi}$.

The various equations have been treated separately so far.
Their consistency follows from showing, using
$
(\varphi^\star D\psi)_\mu
=
\Omega^{-1}\big(\widetilde{D}_\mu\widetilde{\psi}
-\Omega^{-1}\nabla_\mu\Omega\,\widetilde{\psi}\big),
$
that the current $\tilde{\jmath}{}^\mu$ obtained by replacing
$\psi$ and $D_\mu\psi$ by $\widetilde{\psi}$ and
$\widetilde{D}_\mu\widetilde{\psi}$ in Eq.~(2.5),
does verify Eq.~(3.4).
We conclude that $\xi$-{\it preserving conformal transformations
of $M$ are indeed symmetries of the reduced system}.
This generalizes the result known for the ungauged
non-linear Schr\"odinger equation~[8]. Note that
the value $p=1$ in the last term
$|\Psi|^{2p}\Psi$ of (1.1) corresponds to the one dictated by
relativistic conformal invariance of the above mentioned non-linear
Klein-Gordon equation in $n=4$ dimensions.

\goodbreak

\chapter{Conserved quantities}

The traditional form of Noether's theorem uses an action principle
while here we only have field equations. We propose therefore a mixed
approach motivated by Souriau's viewpoint~[9].
Let us start with the `partial' action on Bargmann space $M$
$$
S={1\over2m}\int_M\left\{(D_\mu\psi)^*\,D^\mu\psi
+{R\over 6}\,|\psi|^2-{\lambda\over 2}\,|\psi|^4
\right\}\sqrt{-g}\,d^4\!x,
\equation
$$
with $D_\mu=\nabla_\mu-iea_\mu$ for some vector potential $a_\mu$.

Variation with respect to $\psi^*$ yields the NLWE (2.3).
However, no equation for $a_\mu$ is obtained
because the action (4.1) contains no kinetic term
for the gauge field. We have therefore to {\it add} the FCI (2.1)
as an extra condition.
The variational derivative $-{\delta S/\delta a_\mu}$
yields nevertheless
the current $ej^\mu$ in Eq.~(2.5), which is automatically conserved,
$\nabla_\mu j^\mu=0$, as a consequence of the
invariance of the action $S$ with respect to $U(1)$
gauge transformations.
Similarly, the variational derivative $2\,\delta S/\delta g^{\mu\nu}$
yields, after a tedious calculation,
the energy-momentum tensor $\vartheta_{\mu\nu}$ given by
$$
\eqalign{
3m\,\vartheta_{\mu\nu}
&=(D_\mu\psi)^*\,D_\nu\psi+D_\mu\psi\,(D_\nu\psi)^*
-{1\over 2}\left(\psi^*\,D_\mu D_\nu\psi+
\psi\,(D_\mu D_\nu\psi)^*\right)
\cr
&+{1\over 2}|\psi|^2\left(R_{\mu\nu}-{R\over 6}g_{\mu\nu}\right)
-{1\over 2}g_{\mu\nu}\,(D^\rho\psi)^*\,D_\rho\psi
-{\lambda\over 4}g_{\mu\nu}\,|\psi|^4.
\cr}
\equation
$$
Note the strange coefficient $3$, the Ricci tensor $R_{\mu\nu}$
and the weird term $R/6$.

Using the transformation law of the scalar curvature
and of the covariant derivative, we see that
a conformal transformation $\varphi$ replaces
the integrand in the `partial' action (4.1) by
$$
\left\{(\widetilde{D}_\mu\widetilde{\psi})^*\,
\widetilde{D}^\mu\widetilde{\psi}
+{R\over 6}\,|\widetilde{\psi}|^2-{\lambda\over 2}\,
|\widetilde{\psi|}^4\right\}\sqrt{-g}
-\nabla_\mu\big(|\widetilde{\psi}|^2\,\Omega^{-1}
\nabla^\mu\Omega\big)\sqrt{-g},
\equation
$$
with the notation (3.3),
proving again the conformal symmetry of the NLWE (2.3)
from a more conventional viewpoint.
It follows from the conformal invariance
(or directly from Eqs~(4.2) and (2.3))
that $\vartheta_{\mu\nu}$ is
traceless, $\vartheta^\mu_\mu=0$.
Being the variational derivative with respect to a symmetric tensor,
$\vartheta_{\mu\nu}$ is plainly symmetric.
Finally, the invariance of the
action (4.1) with respect to diffeomorphisms~[9] implies that
$\vartheta_{\mu\nu}$ satisfies the relation
$$
\nabla_\mu\vartheta^{\mu\nu}+ej_\mu f^{\mu\nu}=0,
\equation
$$
where $j^\mu$ is the current found above.
But $j_\mu f^{\mu\nu}$ vanishes because of the FCI (2.1),
yielding
$$
\nabla_\mu\vartheta^{\mu\nu}=0.
\equation
$$
We conclude that $\vartheta_{\mu\nu}$ is a {\it conserved, symmetric
and traceless} energy-momentum tensor.

We now prove Noether's theorem in our framework.
The vector field
$$
k^\mu=\vartheta^\mu_\nu X^\nu,
\equation
$$
($\mu,\nu=0,\ldots,3$), is a conserved current,
$\nabla_\mu k^\mu=0$,
for any conformal vector field $X^\nu$.
Indeed,
$\nabla_\mu(\vartheta^\mu_\nu X^\nu)=
(\nabla_\mu\vartheta^\mu_\nu) X^\nu+
\2\,\vartheta^{\mu\nu}\,L_Xg_{\mu\nu}=0$
using (4.5) and $\vartheta^\mu_\mu=0$.

So far, we have associated conserved four-currents to each
infinitesimal conformal transformation of $(M,g)$.
Let us henceforth assume that $X$ be also $\xi$-preserving and proceed
to derive conserved quantities in $2+1$ dimensions.
Firstly, observe that the current $(k^\mu)$ is $\xi$-invariant
because so is $S$, and hence $\vartheta$, as well as
$X$.
Therefore, it projects into a three-current
$(K^\alpha)=(K^0,\vec{K})$ on ordinary space-time, $Q$.
But $\nabla_s k^s=0$ because $\xi=\partial_s$ is covariantly
constant.
The projected current is thus also conserved,
$$
\nabla_\alpha K^\alpha=0,
\equation
$$
where $\nabla$ is, here, the (non-metric) Newton-Cartan
covariant derivative on $Q$ obtained by projecting the
metric covariant derivative of $(M,g,\xi)$~[3].
The Bargmann metric, $g_{\mu\nu}$, canonically induces a Riemannian
metric, $\gamma_{ij}$, on each $2$-surface $\Sigma_t$ ($t=$const.) of
$Q$.
The quantities
${\cal{Q}}_X=\int_{\Sigma_t}{K^0\sqrt{\gamma}\,d^2\!\vec{x}}$ are
thus conserved provided all currents vanish at infinity.
Now $K^0=\xi_\nu k^\nu$, hence the quantity
$$
{\cal{Q}}_X
=
\int_{\Sigma_t}{
\vartheta_{\mu\nu}X^\mu\xi^\nu\sqrt{\gamma}\,d^2\!\vec{x}},
\equation
$$
does not depend on time $t$ for each $\xi$-preserving
conformal vector field $X$.
The conserved quantities are conveniently calculated
using
$$
\vartheta_{\mu\nu}\xi^\nu
={1\over2i}\left[
\psi^*\,(D_\mu\psi)-\psi\,(D_\mu\psi)^*\right]
-{1\over 6m}\,\xi_\mu\left(
{R\over 6}|\psi|^2
+(D^\nu\psi)^*\,D_\nu\psi
+{\lambda\over2}\,|\psi|^4
\right),
\equation
$$
obtained from Eq.~(4.2) by the equivariance condition (2.4).
In the flat case, for example, a conserved
quantity is associated to each generator in Eq.~(3.1) of
the planar Schr\"odinger group. Using the equivariance and
the field equations we find
$$
\left\{\matrix{
{\cal H}
=
\displaystyle\int\left[\displaystyle{1\over2m}
(\vec{D}\Psi)^*\cdot\vec{D}\Psi
-{\Lambda\over2}(\Psi^*\Psi)^{2}
\right]\,d^2\!\vec{x}
\hfill
&\hbox{energy}\hfill
\ccr
\vec{\cal P}
=
\displaystyle\int\vec{P}\,d^2\!\vec{x}\equiv
\displaystyle\int
\displaystyle{1\over2i}\big[
\Psi^*(\vec{D}\Psi)-(\vec{D}\Psi)^*\Psi\big]\,d^2\!\vec{x}
\qquad
\hfill
&\hbox{momentum}\hfill
\ccr
{\cal J}
=
\displaystyle\int\vec{x}\times\vec{P}\,d^2\!\vec{x}
\hfill
&\hbox{angular momentum}\hfill
\ccr
\vec{\cal G}
=
t\vec{\cal P}
-m\displaystyle\int\vec{x}\,(\Psi^*\Psi)\,d^2\!\vec{x}\hfill
&\hbox{boost}\hfill\ccr
{\cal D}
=
t{\cal H}-\2\,
\displaystyle\int\vec{x}\cdot\vec{\cal P}\,d^2\!\vec{x}
\hfill
&\hbox{dilatation}\hfill
\ccr
{\cal K}
=
t^2{\cal H}-2t{\cal D}
+\displaystyle{m\over2}\,\int r^2(\Psi^*\Psi)\,d^2\!\vec{x}
\hfill
&\hbox{expansion}\hfill
\ccr
{\cal M}
=
m\displaystyle\int\Psi^*\Psi\,d^2\!\vec{x}\hfill
&\hbox{mass}\hfill
\cr
}\right.
\equation
$$
where $\Psi=e^{-ims}\psi$.
Note that the `vortex number'
of Ref.~[1] is associated here
to the vertical Killing vector $\xi$, namely
${\cal M}={\cal Q}_\xi$.

Jackiw and Pi obtain the same conserved quantities by constructing a
non-symmetric energy-momentum tensor $T^{\alpha\beta}$,
($\alpha,\beta=0,1,2$), which satisfies the unusual trace condition
$T^{ij}\delta_{ij}=2T^{00}$. The $T^{00},T^{i0}$ (resp.
$T^{0j},T^{ij}$) components describe the density and flux of the
energy (resp. momentum). Compared to the
$K^0$ and $K^i$ of time (resp. space) translations, we find
$$
\matrix{
&T^{00}
=\displaystyle
-\vartheta^0_0+{1\over6m}\Delta\rho,\hfill
&\hfill
&T^{i0}
=\displaystyle
-\vartheta^i_0-{1\over6m}\partial_i\partial_t\rho,\hfill
\ccr
&T^{0j}
=\displaystyle
\vartheta^0_j,\hfill
&\hfill
&T^{ij}
=\displaystyle
\vartheta^i_j+{1\over3m}
\left(
\delta^i_j\Delta-\partial^i\partial_j
\right)\rho,\hfill
\cr}
\equation
$$
where $\Delta$ denotes the planar Laplace operator and
$\rho=\vert\Psi\vert^2$.
The time components differ in surface terms and yield, therefore, the
same conserved quantities. The lack of symmetry, $T^{0i}\neq T^{i0}$,
is manifest since
$
\vartheta^0_i=\vartheta_{i3}
\neq
-\vartheta_{i0}
-\partial_i\partial_t\rho/6m
$.
However, $T^{ij}=T^{ji}$ because $\vartheta_{\mu\nu}$ is symmetric.
The trace condition of $T^{\alpha\beta}$ comes from the
tracelessness of $\vartheta^\mu_\nu$ and
$\vartheta^3_3=\vartheta^0_0$.
Continuity equations, e.g.
$\partial_\alpha T^{\alpha0}=0$, follow from Eq.~(4.5).

\goodbreak

\chapter{Generalizations}

The FCI (2.1) can easily be generalized by
including a Maxwell-type term,
$$
\sqrt{-g}
\epsilon_{\mu\nu\rho\sigma}\,\xi^\rho\nabla_\tau f^{\tau\sigma}
+\kappa f_{\mu\nu}
=e\sqrt{-g}\,\epsilon_{\mu\nu\rho\sigma}\,\xi^\rho j^\sigma.
\equation
$$
This equation descends, in the same way as before, to the quotient,
$Q$. In the Minkowski case, for example, we still get
$\vec{\nabla}\times\vec{E}+\partial_tB=0$ from
$\partial_{[\mu}f_{\nu\rho]}=0$
while (1.3) is now
generalized to
$$
\kappa B=-e\rho,
\qquad
\partial_iB+\kappa E_i=e\,\epsilon_{ij}\,J^j.
\equation
$$
Note that the equation for $B$ (Gauss' law)
retains the same form it had without
the Maxwell term because $\xi$
is null. The other equation generalizes
Amp\`ere's equation in the
`magnetic-type' Galilean electromagnetism~[10].

Since, in $4$ dimensions, the Maxwell term
$\nabla_\tau{f}^{\tau\sigma}$ goes into
$\Omega^{-4}\nabla_\tau\widetilde{f}^{\tau\sigma}$
under a conformal transformation, the new FCI (5.1) is also invariant
with respect to $\xi$-preserving conformal transformations if the
current changes as in (3.4).
This happens e.g. if we identify the so far unspecified
$j^\mu$ with the right-hand side of Eq.~(2.5). Then Eqs~(2.3--5) and
(5.1) form a coupled system of field equations invariant under the
$\xi$-preserving conformal transformations, i.e. the generalized
Schr\"odinger transformations.

To determine the conserved quantities belonging to these
symmetries we note that Eq.~(5.1) can be written as
$$
\kappa f_{\mu\nu}
=
\sqrt{-g}\,\epsilon_{\mu\nu\rho\sigma}\,\xi^\rho
(ej^\sigma -\nabla_\tau f^{\tau\sigma})
=
\sqrt{-g}\,\epsilon_{\mu\nu\rho\sigma}\,\xi^\rho\hat\jmath{}^\sigma,
\equation
$$
i.e. as Eq.~(2.1) where the new current, $\hat\jmath{}^\mu$,
is the variational derivative
$\hat\jmath{}^\mu=-\delta\hat{S}/\delta{a}_\mu$ of the
action obtained from the expression (4.1) by adding the
standard Maxwell term
$$
\hat{S}=S-{1\over4}\int_M{f_{\mu\nu}f^{\mu\nu}\sqrt{-g}\,d^4\!x}.
\equation
$$
We emphasize that Eq.~(5.1) (or (5.3)) is {\it not} the
variational equation for $a_\mu$ associated with $\hat{S}$ --- which
would require $\hat\jmath{}^\mu$ to vanish --- whereas Eq.~(2.3) is
indeed the variational equation for $\psi^*$.
This shows that the $4$-dimensional dynamics of the $a_\mu$ field we
consider is significantly different from the conventional one.
The variational derivative
$\hat\vartheta_{\mu\nu}=2\,\delta\hat S/\delta g^{\mu\nu}$
yields the energy momentum tensor,
$\hat\vartheta_{\mu\nu}
=
\vartheta_{\mu\nu}+
\Theta_{\mu\nu}$,
where $\vartheta_{\mu\nu}$ is given by Eq.~(4.2) and
$\Theta_{\mu\nu}=f_{\mu\rho}f^\rho_{\ \nu}+{1\over4}\,g_{\mu\nu}
f_{\rho\sigma}f^{\rho\sigma}$ is the standard energy momentum
tensor for the Maxwell field.
Again, $\hat\vartheta_{\mu\nu}$ is a symmetric,
traceless and divergencefree tensor, since all our
previous arguments to establish these properties for
$\vartheta_{\mu\nu}$ apply just as well.
As a consequence, conserved Noether quantities follow immediately
from the expression (4.8) specialized to $\hat\vartheta$.
For example, in the flat case, the only contribution of the Maxwell
field to the nine Noether quantities lies in the
Hamiltonian, $\hat{\cal H}$, since
$
\hat\vartheta_{\mu\nu}\xi^\nu
=
\vartheta_{\mu\nu}\xi^\nu
+{1\over4}\,f_{\rho\sigma}f^{\rho\sigma}\xi_\mu$
due to the fact that
$f_{\mu\nu}\xi^\nu=0$.
In this case we have
$f_{\rho\sigma}f^{\rho\sigma}=2B^2$ and remembering Eq.~(5.2), we
realize that the new Hamiltonian, $\hat{\cal H}$, comes with the
replacement
$
\hat\Lambda=\Lambda+{e^2/\kappa^2}
$
of the coefficient of the quartic term $|\Psi|^4$ in (4.10).

Let us mention for completeness that choosing $\xi$ covariantly
constant and {\it spacelike}, $\xi_\mu\xi^\mu=1$, the quotient of the
Lorentz $4$-manifold $M$ by the integral curves of $\xi$ yields a
Lorentz $3$-manifold of signature $(-,+,+)$. The FCI (2.1) turns out
to project to the {\it same} equation (2.2), explaining why Jackiw
and Pi could use it in their non-relativistic theory. The generalized
FCI (5.1) would in turn project into
$$
\nabla_\alpha F^{\alpha\gamma}
-{\kappa\over2}\sqrt{-g}\,
\epsilon^{\alpha\beta\gamma}F_{\alpha\beta}
=
eJ^\gamma,
\equation
$$
a generalization of the usual curved-space inhomoge\-neous
Maxwell equations. For Min\-kow\-ski space, for example, writing the
metric as $-dt^2+dx^2+dy^2+dz^2$ and choosing $\xi=\partial_z$,
we recover the {\it relativistic} expressions in Ref.~[1], viz
$$
\partial_iE^i+\kappa B=-e\rho,
\qquad
\epsilon_{ij}\partial_tE^j+\partial_iB+\kappa E_i=e\,\epsilon_{ij}J^j,
\equation
$$
to be compared with the non-relativistic expression in Eq.~(5.2).

Reduction of the NLWE (2.3) yields now
a gauged non-linear
Klein-Gordon equation in $2+1$ dimensions for which,
when coupled to the gauge field through a FCI (with or without
Maxwell term),
the $\xi$-preserving conformal transformations still act
as symmetries.
However, no conformal extension of isometries is now obtained as the
$\xi$-preserving conformal transformations merely yield
($\IR$-times) the Poin\-ca\-r\'e group in $2+1$ dimensions.

\goodbreak

\kikezd{Acknowledgements}. We are indebted to Professor
R.~Jackiw for calling our attention to the Schr\"odinger
symmetry of non-relativistic Chern-Simons theory.
One of us (L.~P.) would like to thank Tours University for the
hospitality extended to him
and to the Hungarian National Science and Research Foundation
(Grant No.~$2177$) for a partial financial support.


\vskip1cm


\centerline{\bf\BBig References}

\reference
R.~Jackiw and S-Y.~Pi,
Phys. Rev. Lett. {\bf 64}, 2969 (1990);
Phys. Rev. {\bf D42}, 3500 (1990);
R.~Jackiw,
Ann. Phys. {\bf 201}, 83 (1990); see also
in Proc.
{\it Physics and Mathematics of Anyons}, Houston '91.

\reference
R.~Jackiw, Phys. Today {\bf 25}, 23 (1972);
U.~Niederer, Helv. Phys. Acta {\bf 45}, 802 (1972);
C.~R. Hagen, Phys. Rev. {\bf D5}, 377 (1972);
G.~Burdet and M.~Perrin, Lett. Nuovo Cim. {\bf 4}, 651 (1972).

\reference
C.~Duval, G.~Burdet, H-P.~K\"unzle and M.~Perrin,
Phys. Rev. {\bf D31}, 1841 (1985);
C.~Duval, G.~Gibbons and P.~Horv\'athy,
Phys. Rev. {\bf D43}, 3907 (1991).

\reference
Z.~F.~Ezawa, M.~Hotta and A.~Iwazaki,
Phys. Rev. Lett. {\bf 67}, 411 (1991);
Phys. Rev. {\bf D44}, 452 (1991);
R.~Jackiw and S-Y.~Pi,
Phys. Rev. Lett. {\bf 67}, 415 (1991);
Phys. Rev. {\bf D44}, 2524 (1991).

\reference
K.~Kucha\v r, Phys. Rev. {\bf D22}, 1285 (1980);
C.~Duval and H-P.~K\"unzle, Gen. Rel. Grav. {\bf 16}, 333 (1984).

\reference
G.~Burdet, J.~Patera, M.~Perrin and P.~Winternitz,
J. Math. Phys. {\bf 19}, 1758 (1978).

\reference
V.~De Alfaro, S.~Fubini, P.~Furlan,
Phys. Lett. {\bf 65B}, 163 (1976);
Il Nuovo Cimento {\bf 34}, 569 (1976).

\reference
G.~Burdet and M.~Perrin,
Lett. Math. Phys. {\bf 11}, 147 (1986).

\reference
J.-M.~Souriau, Ann. Inst. H.~Poincar\'e {\bf 20A}, 315 (1974).

\reference
M.~Le Bellac and J.-M.~L\'evy-Leblond,
Il Nuovo Cimento {\bf 14B}, 217 (1973).

\bye